\documentclass[letter]{aa}  

\usepackage{graphicx}
\usepackage{natbib}
\usepackage{hyperref}
\usepackage{txfonts}
\usepackage{color}
\usepackage[scientific-notation=true]{siunitx}
\begin{document} 

\authorrunning{Marchant and Moriya}

   \title{The impact of stellar rotation on the black hole mass-gap from pair-instability supernovae}


   \author{Pablo Marchant
          \inst{1}
          \and
          Takashi J. Moriya\inst{2,3}
          }

   \institute{Institute of Astrophysics, KU Leuven, Celestijnenlaan 200D, 3001, Leuven, Belgium\\
              \email{pablo.marchant@kuleuven.be}
         \and
             National Astronomical Observatory of Japan, National Institutes of Natural Sciences, 2-21-1 Osawa, Mitaka, Tokyo 181-8588, Japan
         \and
         School of Physics and Astronomy, Faculty of Science, Monash University, Clayton, Victoria 3800, Australia
             }


 
  \abstract{Models of pair-instability supernovae (PISNe) predict a gap in black hole (BH) masses between $\sim 45M_\odot-120M_\odot$, which is referred to as the upper BH mass-gap. With the advent of gravitational-wave astrophysics it has become possible to test this prediction, and there is an important associated effort to understand what theoretical uncertainties modify the boundaries of this gap. In this work we study the impact of rotation on the hydrodynamics of PISNe, which leave no compact remnant, as well as the evolution of pulsational-PISNe (PPISNe), which undergo thermonuclear eruptions before forming a compact object. We perform simulations of non-rotating and rapidly-rotating stripped helium stars in a metal poor environment $(Z_\odot/50)$ in order to resolve the lower edge of the upper mass-gap. We find that the outcome of our simulations is dependent on the efficiency of angular momentum transport, with models that include efficient coupling through the Spruit-Tayler dynamo shifting the lower edge of the mass-gap upwards by $\sim 4\%$, while simulations that do not include this effect shift it upwards by $\sim 15\%$. From this, we expect the lower edge of the upper mass-gap to be dependent on BH spin, which can be tested as the number of observed BH mergers increases. Moreover, we show that stars undergoing PPISNe have extended envelopes ($R\sim 10-1000~R_\odot$) at iron-core collapse, making them promising progenitors for ultra-long gamma-ray bursts.}


   \maketitle
%

\section{Introduction}
Very massive stars have long been predicted to undergo pair-instability supernovae (PISNe, \citealt{FowlerHoyle1964,RakaviShaviv1967}) and pulsational-PISNe (PPISNe, \citealt{Fraley1968,Woosley2017}) due to pair-creation in their cores softening the equation of state and inducing instability. Collapse in these conditions leads to runaway oxygen burning and energetic mass ejections. Although there are various candidate electromagnetic transients that could have been powered by this mechanism (cf. \citealt{Gal-Yam+2009,Terreran+2017,Arcavi+2017,Lunnan+2018}) there is no unambiguous event that indicates these transients do occurr in nature.

Indirect evidence for PPISNe/PISNe is provided by gravitational wave observations. PISNe (which leave no remnant) and PPISNe (which result in mass loss before iron-core collapse) have been predicted to result in a gap in black hole (BH) masses between $\sim 45M_\odot-120M_\odot$ \citep{HegerWoosley2002,Yoshida+2016,Woosley2017,Marchant+2019}, which is expected to be an observable feature in the population of binary BH mergers observed by ground base detectors
\citep{Belczynski+2014,Marchant+2016,Belczynski+2016b,SperaMapelli2017}. Results from the first two observing runs of the LIGO and Virgo detectors indicate that there is a dearth of BHs with masses $\gtrsim 45M_\odot$, consistent with the lower edge of the predicted PISNe gap \citep{FishbachHolz2017,LIGOpop}. In the following years additional measurements will further constrain this upper mass gap from PISNe, allowing its use as a standard candle for cosmology \citep{Farr+2019} and as a tool to constrain uncertain nuclear reaction rates \citep{Farmer+2020}.

There is at the moment significant work studying what can impact the predicted location of this mass gap, including uncertainties in nuclear reaction rates \citep{Takahashi2018,Farmer+2019}, convection \citep{Renzo+2020}, the presence of a massive hydrogen envelope \citep{diCarlo+2019}, and accretion after BH formation \citep{vanSon+2020}. Regarding rotation, work has been done to study how it affects the evolution of a star prior to a PPISN/PISN \citep{Chatzopoulos+2013,Mapelli+2020}, but there is still a large uncertainty on how rotation affects the actual hydrodynamics of these events. Early work performed by \citet{Glatzel+1985} shows that rapid rotation can shift the boundaries of instability upwards in mass, but provided no predictions on the resulting properties of BHs formed through this process.

The objective of this letter is to provide a first estimate on how the hydrodynamics of PPISNe/PISNe are affected by rotation, and how this impacts the upper mass gap. In Section \ref{sec:rot} we describe how rotation modifies the criterion for instability. We describe the setup of our numerical simulations of PPISNe and PISNe in Section \ref{sec:methods}, and present our results in Section \ref{sec:results}. We conclude by discussing the implications of our results in Section \ref{sec:discussion}.

\section{Rotation and pair instability}\label{sec:rot}
We model rotation following the shellular approximation, in which all thermodynamical properties of the star are assumed to be constant through rigidly rotating Roche equipotentials. Under this assumption the equations of stellar structure and evolution retain their one dimensional form with rotation being encoded in two coefficients, $f_P$ and $f_T$, that are computed from integrals over the Roche potential \citep{EndalSofia1976,HegerLanger2000}. The momentum equation in this approximation is given by
\begin{eqnarray}
\left(\frac{\partial P}{\partial m_{\Phi}}\right)_t = -\frac{Gm_\Phi}{4\pi r_\Phi^4}f_P-\frac{1}{4\pi r_{\Phi}^2}\left(\frac{\partial r_\Phi}{\partial t}\right)_{m_\Phi},\label{equ:dyn}
\end{eqnarray}
where $r_\Phi$ and $m_\Phi$ represent respectively the volume equivalent radius and the mass associated to each equipotential surface. In the shellular approximation the standard radiative temperature gradient $\nabla_r$ is also scaled by a factor $f_T/f_P$.

The impact of rotation on the hydrodynamics of PPISNe and PISNe can be understood in terms of two different effects produced by centrifugal support: a modification of the stability criterion and that rotating stars follow evolution that resembles that of lower mass stars. 
The variation in the stability criterion can be described by considering a hydrostatic solution of Eq. (\ref{equ:dyn}),
\begin{eqnarray}
\left(\frac{\partial P_0}{\partial m_{\Phi}}\right)_t = -\frac{Gm_\Phi}{4\pi r_{\Phi,0}^4}f_{P,0},\label{equ:dyn2}
\end{eqnarray}
and performing a Lagrangian perturbation on $r_{\Phi,0}$,
\begin{eqnarray}
r_{\Phi}=r_{\Phi,0}+\Delta r_{\Phi}, \quad \Delta r_{\Phi} = \alpha r_{\Phi,0}.
\end{eqnarray}
In terms of the small parameter $|\alpha|\ll1$ the corresponding Lagrangian perturbations for density and pressure are given by
\begin{eqnarray}
\Delta \rho = -3\alpha\rho_0,\quad \Delta P=\frac{\Delta \rho}{\rho}\Gamma_1 P_0=-3\alpha \Gamma_1 P_0,
\end{eqnarray}
where $\Gamma_1\equiv (d\log P/d\log \rho)_{\rm ad}$ is the first adiabatic index of the fluid. As a star contracts or expands its rotation changes, leading to a variation in the $f_P$ correction. From \cite{Paxton+2019} we have that $f_P$ can be approximated as
\begin{eqnarray}
f_{P}=1-\frac{2}{3}\omega^2+\mathcal{O}(\omega^4),\quad
\omega\equiv\frac{\Omega}{\sqrt{GM_\Phi/r_{\rm e}^3}},\label{equ:fp}
\end{eqnarray}
where $\Omega$ is the rotational frequency of a shell and $r_{\rm e}$ is its equatorial radius. The perturbation in $f_{P,0}$ can then be described in terms of the perturbation on $\omega$,
\begin{eqnarray}
\frac{\Delta f_P}{f_{P,0}}=\left(-\frac{4}{3}\omega_0+\mathcal{O}(\omega_0^3)\right)\Delta\omega.
\end{eqnarray}
To compute $\Delta \omega$ we consider that the perturbation preserves the specific angular momentum $j_{\rm rot}$ of each shell. In this case, we have that \citep{Paxton+2019}
\begin{eqnarray}
\frac{j_{\rm rot}}{\sqrt{Gm_\Phi r_\Phi}}=\frac{2}{3}\omega + \mathcal{O}(\omega^3)\rightarrow \Delta\omega=\frac{-\alpha}{2}\omega_0+\mathcal{O}(\omega_0^3).\label{equ:dw}
\end{eqnarray}

Combining Eqs. (\ref{equ:dyn}-\ref{equ:dw}) the acceleration after the perturbation is
\begin{eqnarray}
\frac{1}{4\pi r_{\Phi}^2}\left(\frac{\partial r_\Phi}{\partial t}\right)_{m_\Phi} = \alpha\frac{Gm_\Phi}{4\pi r_{\Phi,0}^4}f_{P,0}\left(4-3\Gamma_1-\frac{2}{3}\omega_0^2 + \mathcal{O}(\omega^4)\right).\label{equ:yay}
\end{eqnarray}
One can now derive a sufficient condition for instability by considering whether or not a contraction of the star, given by $\alpha<0$, leads to runaway collapse. From Eq. (\ref{equ:yay}) we obtain that in order for the fluid to respond to contraction with an inwards acceleration we require that
\begin{eqnarray}
\Gamma_1 < \frac{4}{3}-\frac{2}{9}\omega^2 + \mathcal{O}(\omega^4), \label{equ:gamma}
\end{eqnarray}
which resembles the standard instability criterion $\Gamma_1<4/3$ for a non-rotating self-gravitating body. As a real star does not have a constant $\Gamma_1$ or $\omega$, Eq. (\ref{equ:gamma}) can be true or false in different regions of the star, and whether or not this drives a global instability can be assessed by integrating the difference between the left and right hand sides of the equation through the star \citep{Stothers1999}.

\begin{figure}
   \centering
   \includegraphics[width=\columnwidth]{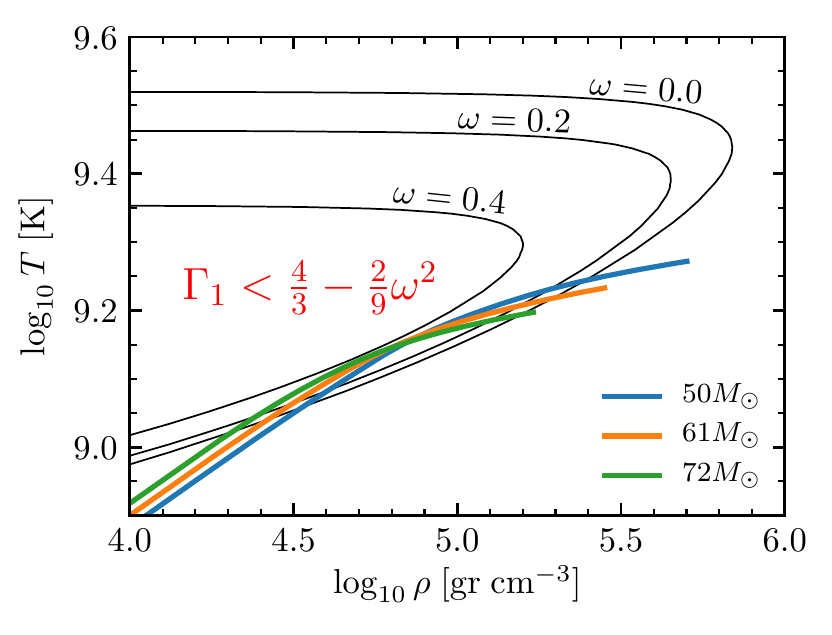}
   \caption{Instability region from pair-creation for different values of the ratio $\omega=\Omega/\sqrt{GM_\Phi/r_e^3}$, computed for material composed of $90\%$ oxygen and $10\%$ carbon by mass. For reference, the profile of three non-rotating stripped star models at the onset of PPISNe/PISNe from \citet{Marchant+2019} are included, with the masses given corresponding to the mass at the onset of instability.}
              \label{fig:gamma}%
\end{figure}

Fig. \ref{fig:gamma} shows in the $\rho-T$ plane the instability region given by Eq. (\ref{equ:gamma}) for different values of $\omega$. Three profiles of non-rotating stellar models at the onset of a PPISN/PISN from \citet{Marchant+2019} are included, with the $50M_\odot$ model undergoing a PPISN and the $61M_\odot$ and $72M_\odot$ models resulting in full disruption through a PISN. For $\omega=0.4$, all three models fall outside of the instability region.

The second effect that can stabilize a rotating star is that its evolution resembles that of a lower mass star. This can be understood in terms of an order of magnitude analysis of Eq. (\ref{equ:dyn}), where if we assume that there is a characteristic value for $f_P$ throughout the star we can estimate the central pressure as
\begin{eqnarray}
\frac{P_c}{M}\sim \frac{GM}{4\pi R^4}f_P,
\end{eqnarray}
where $R$ and $M$ are the radius and mass of the star. Using Eq. (\ref{equ:fp}), taking $\rho_c\propto M/R^3$ and assuming a gas strongly dominated by radiation pressure such that $P\propto T^4$, we find that
\begin{eqnarray}
\frac{T_c^3}{\rho_c}\propto M^{1/2}(1-2\omega^2/3)^{3/4}.
\end{eqnarray}
What this implies is that more massive stars follow an evolution in the $\rho_c-T_c$ plane at higher $T_c$ for a given $\rho_c$, thus approaching the region where $\Gamma_1<4/3$. Rotation lowers the value of the central temperature at a fixed central density, causing the star to evolve further away from the instability region.

\section{Methods}\label{sec:methods}
We perform our numerical simulations using version 13311 of the \texttt{MESA} code for stellar structure and evolution \citep{Paxton+2011,Paxton+2013,Paxton+2015,Paxton+2018,Paxton+2019}, with the setup described in \citet{Marchant+2019}. A detailed description of our simulation setup is provided in Appendix \ref{app:mesa}.

As initial conditions we use pure helium stars, which are representative of binary BH progenitors formed through the isolated evolution of close binaries. To maximize the angular momentum content of our models at the onset of PPISNe/PISNe, we consider helium stars at a metallicity of $Z_\odot/50$ with $Z_\odot=0.0142$ \citep{Asplund+2009}. One important process we take into account is the Spruit-Tayler (ST) dynamo for angular momentum transport \citep{Spruit1999,Spruit2002}. The inclusion of the ST dynamo in stripped stars leads to near solid body rotation and efficient loss of angular momentum from winds. In particular, \citet{Qin+2019} showed that binary models without the ST dynamo can reproduce the near critical spins of BHs observed in high-mass X-ray binaries, while models that include it result in BHs with near zero spin. The physical nature of the ST dynamo is currently a topic of active discussion (cf. \citealt{DenissenkovPinsonneault2007,Zahn+2007,Fuller+2019}), so we consider models with and without this mechanism.

Intial rotation rates are set in our simulations as solid body rotation at the beginning of core helium burning. The angular frequency is taken to be $90\%$ of its critical value at the surface $\Omega_{\rm crit}$ which is given by \citep{Langer1997}
\begin{eqnarray}
\Omega_c = \sqrt{\frac{GM(\Gamma-1)}{R_e^3}},
\end{eqnarray}
where $R_e$ is the equatorial radius of the star and the Eddington factor $\Gamma$ is defined as
\begin{eqnarray}
\Gamma\equiv \frac{\kappa L}{4\pi c G M}.
\end{eqnarray}
For comparison, we also compute non-rotating models. The initial masses in our simulations are chosen to cover the range of masses at which PPISNe occur, while resolving the boundaries between non-pulsating and pulsating models and between pulsating and fully disrupted models. We model evolution until either the star is completely disrupted in a PISN, or an iron-core is formed and collapses. Tables summarizing each individual simulation we performed are included in Appendix \ref{app:tables}.

\section{Results}\label{sec:results}
\begin{figure}
   \centering
   \includegraphics[width=\columnwidth]{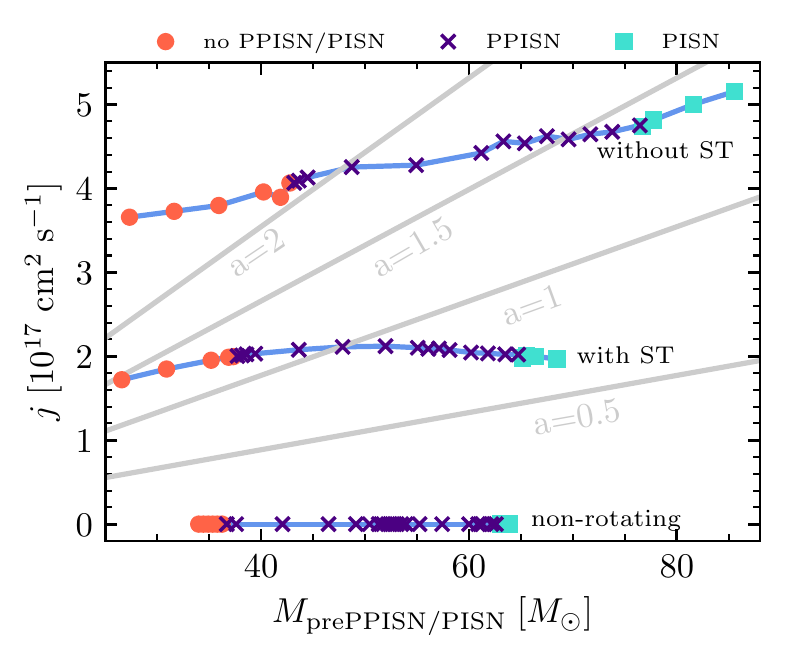}
   \caption{Specific angular momentum $j$ and masses at the onset of PPISNe/PISNe for all of our simulations. For models that do not undergo pair-instability values correspond to the moment of iron-core collapse. Gray lines correspond to constant values of the spin parameter $a=jc/MG$.}
              \label{fig:grid}%
\end{figure}

The properties of our models at the onset of pair-instability are illustrated in Fig. \ref{fig:grid}, indicating as well the outcome of the simulations in terms of the occurrence of PPISNe or PISNe. We define the specific angular momentum $j$ as the total angular momentum of the star divided by its mass. As expected, models that do not include the ST dynamo retain more angular momentum. We find that all our models with the ST dynamo evolve towards critical rotation at their surface ($\Omega/\Omega_c=1$) during the contraction phase between core helium depletion and core carbon ignition. To prevent models from evolving above critical rotation we consider enhanced wind mass loss as described in \citet{Paxton+2015}, such that the star removes sufficient angular momentum to remain below critical. This results in enhanced mass-loss at this late stage \citep{Aguilera-Dena+2018}. As the models with the ST dynamo evolve as solid body rotators even at these late phases, these simulations represent an upper limit on the angular momentum content at the onset of pair-instability if such stars have strong angular momentum coupling.

Our non-rotating models can be used as a baseline to assess the impact of rotation. Without rotation we find that PPISNe occur for masses between $36.7M_\odot-62.6M_\odot$, with models below this range evolving hydrostatically until iron-core collapse and models above being disrupted in a PISN. Rotating models with the ST dynamo shift this range upwards to $37.7M_\odot-64.7M_\odot$, while the range is between $43.2M_\odot-76.5M_\odot$ when the ST dynamo is not included. Thus, between our non-rotating and rotating simulations we find a $\sim 20\%$ shift in the mass range for the onset of PPISNe.

\begin{figure}
   \centering
   \includegraphics[width=\columnwidth]{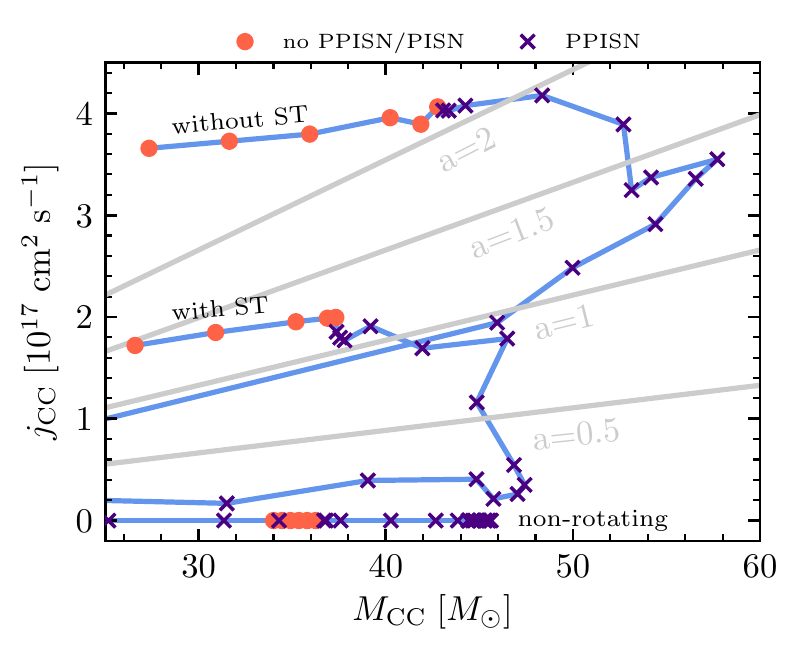}
   \caption{Same as Fig. \ref{fig:grid} but for the properties at the onset of iron-core collapse. $M_{\rm CC}$ corresponds to the baryonic mass of the star at core-collapse.}
              \label{fig:CC}%
\end{figure}

Fig. \ref{fig:CC} summarizes the masses and angular momentum of our models at the point of iron-core collapse. Of particular interest is the reduction of the spin parameter $a=jc/MG$ at core-collapse compared to that at the onset of PPISNe. For example, in our simulations with the ST dynamo the most massive model at core-collapse has $47.4M_\odot$ and a spin of $0.17$, while at the onset of the PPISN it had $56.1M_\odot$ and a spin of $0.84$. This large reduction in spin is caused not only by mass loss but is also due to angular momentum transport between a compact core and an extended envelope. As shown by \citet{Marchant+2019}, heat injected by the thermonuclear pulses leads to a quiescent phase lasting up to ten thousand years where the outer layers of the star can expand beyond $100R_\odot$. Our rotating simulations with the ST dynamo transport angular momentum efficiently to these extended layers which are ejected in later pulses. In contrast, simulations without the ST dynamo do not undergo efficient angular momentum transport during this phase; the model with the highest mass at core-collapse has $57.7M_\odot$ and a spin of $1.39$, and corresponds to a pre-PPISN star of $63.7M_\odot$ with a spin of $1.57$.

Many of our simulations at core-collapse have spin parameters in excess of unity, such that assuming direct collapse to a BH is not an adequate model. Instead, we use the model of \citet{BattaRamirezruiz2019} by assuming that the innermost $3M_\odot$ of the star collapses to a BH with a maximum spin parameter of unity. The remainder of the star falls directly into the BH or is accreted through a disk, releasing energy and angular momentum in such a way that the resulting BH has $a\leq1$. In determining the final gravitational mass of the BH we ignore energy losses from neutrino emission. This is justified since even if the star undergoes collapse to a BH via an intermediate proto-neutron star phase, this would only reduce by $\sim 10\%$ the gravitational mass of the collapsing iron-core which is itself just $\sim 10\%$ of the total mass of the star (cf. appendix A.2 of \citealt{Zevin+2020}).

\begin{figure}
   \centering
   \includegraphics[width=\columnwidth]{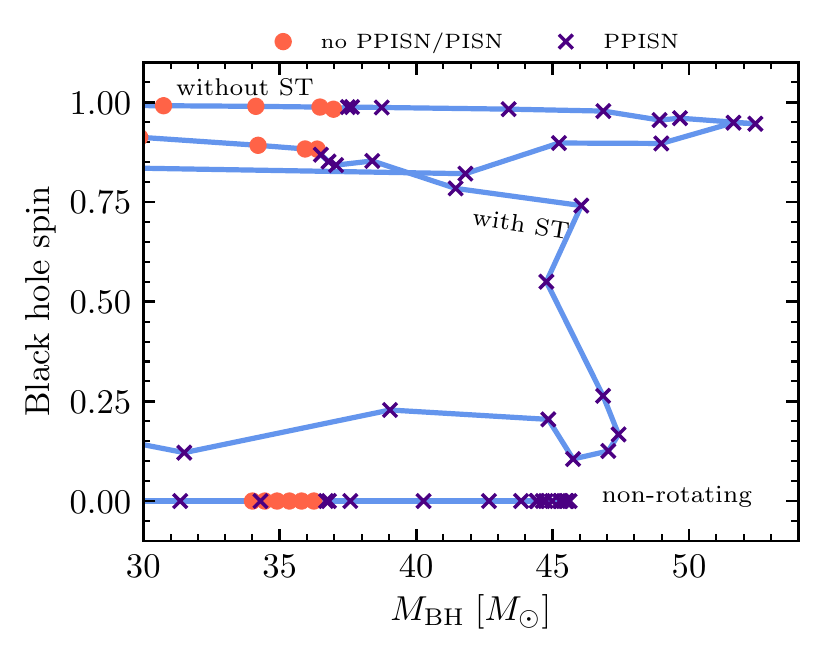}
   \caption{Final gravitational BH masses and spins predicted from the properties of our simulations at iron core-collapse combined with the model of \citet{BattaRamirezruiz2019} for BH formation.}
              \label{fig:BH}%
\end{figure}

The resulting gravitational mass of the BHs formed, together with their spins, are shown in Fig. \ref{fig:BH}. For non-rotating models the lower edge of the PISNe mass-gap is at $45.5M_\odot$, while for rotating models with and without the inclusion of the ST dynamo the maximum masses we obtain are $47.4M_\odot$ and $52.4M_\odot$ respectively. This represents an upwards shift in the mass at the edge of the gap of $\sim 4\%$ and $\sim 15\%$ for the cases with and without the ST dynamo. 

\section{Discussion}\label{sec:discussion}
In order to study how rotation impacts the lower edge of the BH mass gap predicted from PISNe, we have performed simulations of PPISNe and PISNe from rapidly rotating helium star progenitors by using a 1D approximation for hydrodynamics. We find that the final outcome of our simulations depends on the strength of angular momentum transport. Compared to non-rotating models, rotating simulations that include strong coupling via the ST dynamo produce a small increase ($\sim 4\%$) on the mass range at which PPISNe occur, as well as on the final masses of the BHs produced. The effect is larger in simulations without the ST dynamo, with an increase of $\sim 15\%$ in the mass range for PPISNe and its resulting BH masses. This points to the lower edge of the PISNe mass gap increasing in mass at higher BH spins ($a\gtrsim 0.8$), as shown in Fig. \ref{fig:BH}. Assessing this prediction with observations of merging binary BHs presents important complications. In most cases only the effective spin $\chi_{\rm eff}$ rather than the individual BH spins can be measured to any accuracy, and there is a significant degeneracy with respect to the mass ratio of the system \citep{Hannam+2013}. Additionaly, BHs produced in binary BH mergers can also produce a population of high-spin and high mass BHs in the upper mass gap \citep{GerosaBerti2017}. 

Despite these uncertainties one particular object of interest in the first catalogue of gravitational wave transients is GW170729 \citep{GWTC1}, for which the effective spin was reported to be $\chi_{\rm eff}=0.37^{+0.21}_{-0.25}$ with a mass of the primary BH of $50.2^{+16.2}_{-10.2}M_\odot$. Although the mass of the primary BH in GW170729 is consistent with the edge of the mass gap as predicted by non-rotating models, most of the $90\%$ credible interval falls within the gap, which has motivated discussions on GW170729 being a second generation BH merger \citep{Kimball+2020}. Analyzing the posterior distributions provided by the LIGO-Virgo collaboration we find that the individual spin of this black hole is $a=0.69^{+0.27}_{-0.55}$ (see appendix \ref{app:gw170729}), making it a potential candidate for a BH formed through a rapidly rotating star that underwent PPISN. Even with imperfect measurements, a large number of detections can be used to derive the intrinsic properties of the population \citep{Mandel+2019}, which will provide stronger evidence than inferences based on individual objects.

One additional aspect that we have not considered here is the relevance of our simulations in the context of long gamma-ray bursts (LGRBs) progenitors. The standard model for LGRBs invokes a so-called collapsar, where the collapse of a star with a sufficiently high angular momentum can lead to the formation of a massive disk around a newly formed BH \citep{Woosley1993, MacfadyenWoosley1999}. The stellar origin of LGRBs is supported by the observation of associated SNe (cf. \citealt{Galama+1998}), with the lack of hydrogen and helium in the spectra of these SNe pointing to stripped stars as the progenitors (cf. \citealt{Campana+2006}).

\begin{figure}
   \centering
   \includegraphics[width=\columnwidth]{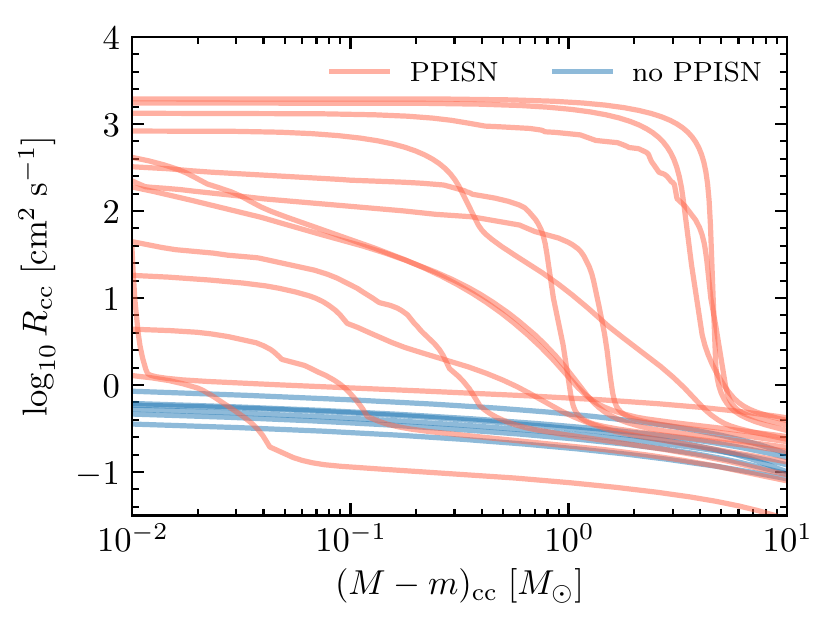}
   \caption{Radii as a function of outer mass coordinate for all our simulations without the ST dynamo at the point of iron-core collapse. The total mass $M$ is defined as the innermost mass coordinate for which the velocity is below the local escape velocity $v_{\rm esc}=\sqrt{2Gm/r}$.}
              \label{fig:radius}%
\end{figure}

From our simulations of rotating stripped stars we can study whether the occurrence of pulsational mass loss prior to collapse can have an impact on a potential LGRB or its associated SNe. As shown by \citet{Marchant+2019} strong pulses are expected to deposit energy throughout the layers of the star that remain bound, leading to an expansion of the star by orders of magnitude prior to iron-core collapse. This is illustrated in Fig. \ref{fig:radius}, where we plot the radius at core-collapse of our models computed without the ST-dynamo versus the mass coordinate measured inwards from the surface. Models that do not pulsate have radii $\lesssim 1R_\odot$, while pulsating models cover a large range of radii going well beyond $10 R_\odot$. If all these collapsing models resulted in LGRBs, the large variety of free-fall timescales for these extended envelopes can potentially translate to different LGRB durations, reaching into the regime of ultra-long GRBs \citep[e.g.,][]{levan2014ulgrb}. The free-fall timescales of the progenitors with radii above $10 R_\odot$ exceeds $10^4~\mathrm{sec}$ and corresponds to those observed for ultra-long GRBs ($\sim10^4~\mathrm{sec}$). Many progenitors here have free-fall timescales exceeding $10^5~\mathrm{sec}$, which are well beyond the duration of ultra-long GRBs, but the accretion could be suppressed at some moment due to, e.g., accompanying SNe. There also exists a candidate GRB with a duration of the order of $10^7~\mathrm{sec}$ \citep{quataert2012} and it matches the free-fall timescales of the most extended progenitors ($\sim 10^3 R_\odot$) produced by PPISNe. In the accompanying paper \citep{mmb20}, we investigate the explosion properties of one extended GRB progenitor presented here and show that it can also explain the peculiar SN component associated with the ultra-long GRB~111209A \citep{greiner2015}.

As already mentioned the objective of this study is to provide a first analysis of the impact of rotation on the hydrodynamic evolution of stars undergoing PPISNe, but an important caveat needs to be pointed out. The use of 1D simulations is dependent on the shellular approximation, but during rapid hydrodynamical phases horizontal turbulence, which is the process believed to produce near-shellular rotation in radiative layers of a rotating star \citep{ChaboyerZahn1992}, cannot operate fast enough. This implies that our simulations can be used to determine the masses for which PPISNe/PISNe occurr, but there is still an important quantitative uncertainty on final BH masses produced by these thermonuclear events. Nevertheless, our results can serve to study the growing sample of gravitational wave sources with non-negligible spins, while motivating multi-D calculations of PPISNe/PISNe.

\begin{acknowledgements}
       PM acknowledges support from the FWO junior postdoctoral fellowship No. 12ZY520N.
       TJM is supported by the Grants-in-Aid for Scientific Research of the Japan Society for the Promotion of Science (JP18K13585, JP20H00174).
\end{acknowledgements}
\bibliographystyle{aa}

\appendix

\section{Details of \texttt{MESA} simulations} \label{app:mesa}
In this appendix we briefly summarize the physical assumptions and ingredients used in our simulations. \texttt{MESA} uses an equation of state that is constructed from a patchwork of results that have different ranges of validity, including OPAL \citep{RogersNayfonov2002}, SCVH \citep{Saumon+1995}, PC \citep{PotekhinChabrier2010} and HELM \citep{TimmesSwesty2000}. Opacities are computed using tables from the OPAL project \citep{IglesiasRogers1996}, with metal abundances corresponding to scaled solar values as measured by \citet{Asplund+2009}. At low temperatures we rely on the opacity tables of \citet{Ferguson+2005}\footnote{Note that these low temperature tables are for hydrogen rich material which does not correspond to the case in our simulations. These low temperature are reached on the outermost layers of our models that expand to large radii ($R>10^3R_\odot$), but do not affect the conclusion that these stars would expand to such a large size after pulsations.}. Nuclear reaction rates are taken from \citet{Angulo+1999} and \citet{CaughlanFowler1988} with a preference for the former when available. During PPISNe/PISNe we make use of the nuclear network \texttt{approx\_21\_plus\_co56.net}, which is the same 21 isotope network described in \citet{Marchant+2019} with the inclusion of $^{56}$Co to better account for the radioactive decay of $^{56}$Ni. Mass loss rates are computed as a combination of the prescriptions of \citet{Vink+2001}, \citet{Hamann+1995} and \citet{NieuwenhuijzendeJager1990}, as described in \citet{Marchant+2019}. We model convection using mixing-length theory \citep{Bohm-Vitense1958,CoxGiuli1968} and a mixing length parameter $\alpha_{\rm MLT}=2$ with exponential overshooting at convective boundaries \citep{Herwig2000} defined by parameters $f=0.01$ and $f_0=0.005$. Semiconvective mixing is modelled as as in \citet{Langer+1983} with an efficiency parameter $\alpha_{\rm sc}=1$.

Hydrodynamical evolution is computed using the HLLC Riemann solver developed by \citet{Toro+1994}, with gravity being scaled by the $f_P$ parameter as described in Eq. \ref{equ:dyn}. The value of $f_P$ is computed following \citet{Paxton+2019}. In addition to the ST dynamo, we include angular momentum transport in our simulations from Eddington-Sweet circulations, the GSF instability and both secular and dynamical shear following the method of \citet{Heger+2000}. As there can be long periods of quiescence between events of pulsational mass loss and iron-core collapse, if a star restores hydrostatic equilibrium after a pulse we remove the ejected material from our simulation grid by following the method described in \citet{Paxton+2018} and \citet{Marchant+2019}.

All necessary input files to reproduce our simulations, as well as machine readable tables with our results, are available for download at \url{https://doi.org/10.5281/zenodo.3940339}. 

\section{Tabulated results}\label{app:tables}
The results of our simulations are summarized in Tables \ref{table:norot}
, \ref{table:rotST} and \ref{table:rotnoST} for our models that are non-rotating, rotating with the ST dynamo, and rotating without the ST dynamo respectively. The properties listed are
\begin{itemize}
\item $M_{\rm i}$: Initial mass of the helium star.
\item $M_{\rm He\;dep}$: Mass of the star at core-helium depletion.
\item $M_{\rm CO,\;He\;dep}$: Mass of the carbon-oxygen core of the star at core helium depletion, defined as the innermost mass boundary where the mass fraction of helium is below $1\%$.
\item $M_{\rm pre\;PPISN/PISN}$: Mass at the onset of the PPISN/PISN.
\item $M_{\rm ejecta}$: Mass ejected through pulsations. For models undergoing a PISN this is equal to $M_{\rm pre\;PPISN/PISN}$.
\item $M_{\rm CC}$: Baryonic mass of the star at iron-core collapse. Note that $M_{\rm CC}\neq M_{\rm pre\;PPISN/PISN}+M_{\rm ejecta}$ as wind mass loss during quiescent periods between pulsations and iron-core collapse can contribute.
\item \# of pulses: Number of mass ejections produced by a PPISN/PISN.
\item Duration: Time between the onset of the PPISN until iron-core collapse
\item max KE: Maximum kinetic energy of ejected material achieved in an individual pulse.
\item $a_{\rm i}$, $a_{\rm He\;dep}$, $a_{\rm pre\;PPISN/PISN}$ and $a_{\rm CC}$: Spin parameter $a=jc/MG$ for the layers of the star that are below the escape velocity at different phases.
\item $M_{\rm BH}$, $a_{\rm BH}$: Final gravitational mass and spin of the BH.
\end{itemize}
\begin{table*}
\caption{Summary of results for non-rotating models.}\label{table:norot}
\centering
\begin{tabular}{ccccccccccccccc} 
\hline\hline 
$M_{\rm i}$ &$M_{\rm He\;dep}$ &$M_{\rm CO,\;He\;dep}$ &$M_{\rm pre\;PPISN/PISN}$ &$M_{\rm ejecta}$ &$M_{\rm CC}$ &\# of pulses &Duration &max KE\\ 
$(M_{\odot})$ &$(M_{\odot})$ &$(M_{\odot})$ &$(M_{\odot})$ &$(M_{\odot})$ &$(M_{\odot})$ & &(yr) &$10^{51} [erg]$\\ 
\hline
\num{36.0000} & \num{34.03213075} & \num{29.8009469} & - & - & \num{34.00064421} & 0 & - & -\\ 
\num{36.5000} & \num{34.48450498} & \num{30.25625091} & - & - & \num{34.45240902} & 0 & - & -\\ 
\num{37.0000} & \num{34.93465392} & \num{30.58656165} & - & - & \num{34.90182163} & 0 & - & -\\ 
\num{37.5000} & \num{35.38786561} & \num{31.08518929} & - & - & \num{35.35434911} & 0 & - & -\\ 
\num{38.0000} & \num{35.83481139} & \num{31.43489027} & - & - & \num{35.80070478} & 0 & - & -\\ 
\num{38.5000} & \num{36.28658188} & \num{31.89229239} & - & - & \num{36.2516551} & 0 & - & -\\ 
\num{39.0000} & \num{36.73381489} & \num{32.23988931} & \num{36.6982931} & \num{1.07154903e-05} & \num{36.69828238} & 1 & \num[scientific-notation=true]{0.0001422099303} & \num[scientific-notation=true]{1.36364565e-06}\\ 
\num{40.0000} & \num{37.63666687} & \num{33.13484924} & \num{37.59980067} & \num{0.01437260045} & \num{37.58542807} & 1 & \num[scientific-notation=true]{0.0003751598415} & \num[scientific-notation=true]{0.0006197922965}\\ 
\num{45.0000} & \num{42.10992527} & \num{37.22552796} & \num{42.06600327} & \num{1.793675941} & \num{40.27232733} & 5 & \num[scientific-notation=true]{0.01299976447} & \num[scientific-notation=true]{0.09646649978}\\ 
\num{50.0000} & \num{46.54057325} & \num{41.31886361} & \num{46.48887332} & \num{1.850637715} & \num{44.6382356} & 5 & \num[scientific-notation=true]{0.08758408256} & \num[scientific-notation=true]{0.04488535174}\\ 
\num{53.0000} & \num{49.18562623} & \num{43.72189919} & \num{49.12950854} & \num{4.143203196} & \num{44.98630534} & 9 & \num[scientific-notation=true]{1.391288337} & \num[scientific-notation=true]{0.1120804028}\\ 
\num{54.5000} & \num{50.50570146} & \num{44.89635163} & \num{50.44705559} & \num{4.823114349} & \num{45.62393403} & 6 & \num[scientific-notation=true]{5.504435624} & \num[scientific-notation=true]{0.1466935056}\\ 
\num{55.5000} & \num{51.38255396} & \num{45.7201379} & \num{51.32232206} & \num{6.01110562} & \num{45.31092313} & 5 & \num[scientific-notation=true]{24.02052338} & \num[scientific-notation=true]{0.2023289595}\\ 
\num{56.0000} & \num{51.81855924} & \num{46.00163243} & \num{51.75797554} & \num{6.481084627} & \num{45.27565846} & 5 & \num[scientific-notation=true]{70.41699336} & \num[scientific-notation=true]{0.2437843148}\\ 
\num{56.5000} & \num{52.25705934} & \num{46.51120409} & \num{52.19521868} & \num{6.712026517} & \num{45.48111735} & 5 & \num[scientific-notation=true]{104.4261517} & \num[scientific-notation=true]{0.2595803592}\\ 
\num{57.0000} & \num{52.69616542} & \num{46.92280766} & \num{52.63383814} & \num{7.104630252} & \num{45.5238489} & 4 & \num[scientific-notation=true]{232.1935937} & \num[scientific-notation=true]{0.3059412726}\\ 
\num{57.5000} & \num{53.12837281} & \num{47.26746583} & \num{53.0654592} & \num{8.30272175} & \num{44.74725657} & 4 & \num[scientific-notation=true]{582.8770715} & \num[scientific-notation=true]{0.3767231718}\\ 
\num{58.0000} & \num{53.56744626} & \num{47.64454788} & \num{53.50367781} & \num{9.039810086} & \num{44.43938311} & 4 & \num[scientific-notation=true]{890.9471595} & \num[scientific-notation=true]{0.4293070256}\\ 
\num{58.5000} & \num{54.00583099} & \num{48.08950871} & \num{53.94120725} & \num{9.494571599} & \num{44.41204532} & 4 & \num[scientific-notation=true]{1229.306248} & \num[scientific-notation=true]{0.47888019}\\ 
\num{60.0000} & \num{55.31379782} & \num{49.28746332} & \num{55.24671327} & \num{11.3324948} & \num{43.84655514} & 4 & \num[scientific-notation=true]{2277.665432} & \num[scientific-notation=true]{0.6638091969}\\ 
\num{62.5000} & \num{57.48881771} & \num{51.30453501} & \num{57.41723523} & \num{14.63535475} & \num{42.6690728} & 3 & \num[scientific-notation=true]{3913.016102} & \num[scientific-notation=true]{1.102343688}\\ 
\num{65.5000} & \num{60.08709056} & \num{53.67073506} & \num{60.01066271} & \num{23.08130361} & \num{36.80346349} & 4 & \num[scientific-notation=true]{5901.827543} & \num[scientific-notation=true]{2.095663707}\\ 
\num{66.5000} & \num{60.95111695} & \num{54.47047657} & \num{60.87295612} & \num{26.46562826} & \num{34.29470216} & 5 & \num[scientific-notation=true]{6522.547123} & \num[scientific-notation=true]{2.534913622}\\ 
\num{67.0000} & \num{61.38191925} & \num{54.76077812} & \num{61.30292497} & \num{29.85331478} & \num{31.35189251} & 8 & \num[scientific-notation=true]{6872.510981} & \num[scientific-notation=true]{2.785815875}\\ 
\num{68.0000} & \num{62.24293857} & \num{55.63872256} & \num{62.16263728} & \num{36.94621093} & \num{25.17431048} & 2 & \num[scientific-notation=true]{8224.975101} & \num[scientific-notation=true]{3.374353613}\\ 
\num{68.5000} & \num{62.67600605} & \num{55.94584537} & \num{62.5943631} & \num{45.58076593} & \num{16.99374307} & 1 & \num[scientific-notation=true]{10360.99164} & \num[scientific-notation=true]{3.655189731}\\ 
\num{69.0000} & \num{63.10715935} & \num{56.43643545} & \num{63.02450171} & \num{63.02450171} & - & 1 & - & \num[scientific-notation=true]{3.889812225}\\ 
\num{69.5000} & \num{63.53693624} & \num{56.80409074} & \num{63.45400531} & \num{63.45400531} & - & 1 & - & \num[scientific-notation=true]{4.197065357}\\ 
\num{70.0000} & \num{63.96747304} & \num{57.17093687} & \num{63.8831794} & \num{63.8831794} & - & 1 & - & \num[scientific-notation=true]{4.519273375}\\ 
\hline
\end{tabular}
\end{table*}

\begin{sidewaystable*}
\caption{Summary of results for rotating models including angular momentum transport via the ST dynamo.}\label{table:rotST}
\centering
\begin{tabular}{ccccccccccccccc} 
\hline\hline           
$M_{\rm i}$ &$M_{\rm He\;dep}$ &$M_{\rm CO,\;He\;dep}$ &$M_{\rm pre\;PPISN/PISN}$ &$M_{\rm ejecta}$ &$M_{\rm CC}$ &\# of pulses &Duration &max KE &$a_{\rm i}$ &$a_{\rm He\;dep}$ &$a_{\rm pre\;PPISN/PISN}$ &$a_{\rm CC}$ &$M_{\rm BH}$ &$a_{\rm BH}$\\ 
$(M_{\odot})$ &$(M_{\odot})$ &$(M_{\odot})$ &$(M_{\odot})$ &$(M_{\odot})$ &$(M_{\odot})$ & &(yr) &$10^{51} [erg]$ & & & & &$(M_{\odot})$ &\\ 
\hline
\num{30.0000} & \num{27.16925984} & \num{23.06570992} & - & - & \num{26.60009985} & 0 & - & - & \num{5.810768685} & \num{2.314598175} & \num{1.459773915} & \num{1.459773915} & \num{25.54669952} & \num{0.9297704731}\\ 
\num{35.0000} & \num{31.43920497} & \num{27.18216082} & - & - & \num{30.90565099} & 0 & - & - & \num{5.301237356} & \num{1.979649742} & \num{1.349701428} & \num{1.349701428} & \num{29.86873117} & \num{0.9123763126}\\ 
\num{40.0000} & \num{35.68247573} & \num{31.34100477} & - & - & \num{35.19679603} & 0 & - & - & \num{4.843385051} & \num{1.713472074} & \num{1.252498735} & \num{1.252498735} & \num{34.2059979} & \num{0.8921505111}\\ 
\num{42.0000} & \num{37.34848597} & \num{32.64514651} & - & - & \num{36.8817461} & 0 & - & - & \num{4.717212686} & \num{1.625012312} & \num{1.216934216} & \num{1.216934216} & \num{35.93229794} & \num{0.8829610926}\\ 
\num{42.5000} & \num{37.79155683} & \num{32.94746677} & - & - & \num{37.32991048} & 0 & - & - & \num{4.654815921} & \num{1.602705646} & \num{1.206740122} & \num{1.206740122} & \num{36.37129836} & \num{0.8825900186}\\ 
\num{43.0000} & \num{38.19073336} & \num{33.25950425} & \num{37.73090074} & \num{0.3626225403} & \num{37.3682782} & 1 & \num[scientific-notation=true]{0.001559396915} & \num[scientific-notation=true]{0.02716051505} & \num{4.63996839} & \num{1.582912464} & \num{1.198064418} & \num{1.120194069} & \num{36.50854541} & \num{0.8685375695}\\ 
\num{43.5000} & \num{38.62975976} & \num{33.75319586} & \num{38.17883881} & \num{0.6198045096} & \num{37.5590343} & 1 & \num[scientific-notation=true]{0.00363759126} & \num[scientific-notation=true]{0.05194061776} & \num{4.579484288} & \num{1.560562846} & \num{1.188381799} & \num{1.080176557} & \num{36.78926252} & \num{0.8513718422}\\ 
\num{44.0000} & \num{39.04421435} & \num{34.04847855} & \num{38.59210217} & \num{0.7978906559} & \num{37.79421152} & 1 & \num[scientific-notation=true]{0.005904700665} & \num[scientific-notation=true]{0.05988986613} & \num{4.57547197} & \num{1.553324645} & \num{1.183035458} & \num{1.057364051} & \num{37.06538707} & \num{0.8424762247}\\ 
\num{45.0000} & \num{39.88336233} & \num{34.85400694} & \num{39.44761729} & \num{0.2638861146} & \num{39.18373118} & 1 & \num[scientific-notation=true]{0.003207965463} & \num[scientific-notation=true]{0.01192604331} & \num{4.480525995} & \num{1.501168098} & \num{1.161946217} & \num{1.099481953} & \num{38.39294261} & \num{0.8529286934}\\ 
\num{50.0000} & \num{44.02579848} & \num{38.6208392} & \num{43.63518174} & \num{1.679755575} & \num{41.95542617} & 3 & \num[scientific-notation=true]{0.02112068859} & \num[scientific-notation=true]{0.07446376723} & \num{4.193686508} & \num{1.326803493} & \num{1.0747861} & \num{0.9107346506} & \num{41.44696996} & \num{0.7840809887}\\ 
\num{55.0000} & \num{48.18016613} & \num{42.46345741} & \num{47.84576714} & \num{1.355523229} & \num{46.49024391} & 9 & \num[scientific-notation=true]{0.03208843386} & \num[scientific-notation=true]{0.06829913221} & \num{3.90015028} & \num{1.177693142} & \num{0.9963695146} & \num{0.867926724} & \num{46.05326512} & \num{0.7409890165}\\ 
\num{60.0000} & \num{52.26799603} & \num{46.23852222} & \num{51.9735687} & \num{7.113532194} & \num{44.86003651} & 9 & \num[scientific-notation=true]{1.261092685} & \num[scientific-notation=true]{0.407103016} & \num{3.686956103} & \num{1.055745411} & \num{0.9209913702} & \num{0.5839018929} & \num{44.77275422} & \num{0.5501333074}\\ 
\num{63.7500} & \num{55.34281567} & \num{49.21900745} & \num{55.07222008} & \num{8.214213801} & \num{46.85036506} & 4 & \num[scientific-notation=true]{296.8916756} & \num[scientific-notation=true]{0.2760686066} & \num{3.513731943} & \num{0.9704304754} & \num{0.8614449516} & \num{0.2629400362} & \num{46.8499295} & \num{0.2638607572}\\ 
\num{65.0000} & \num{56.35696274} & \num{50.04538834} & \num{56.09079105} & \num{8.642381215} & \num{47.42023575} & 3 & \num[scientific-notation=true]{944.8372468} & \num[scientific-notation=true]{0.3859729598} & \num{3.466730137} & \num{0.9437285839} & \num{0.8407538048} & \num{0.1669159633} & \num{47.41847931} & \num{0.1670841894}\\ 
\num{66.2500} & \num{57.37576968} & \num{51.03147418} & \num{57.12017238} & \num{10.02155461} & \num{47.0423212} & 3 & \num[scientific-notation=true]{1787.202635} & \num[scientific-notation=true]{0.556669073} & \num{3.418178098} & \num{0.9218022441} & \num{0.8272699127} & \num{0.1251508941} & \num{47.04218898} & \num{0.125505265}\\ 
\num{67.5000} & \num{58.3871185} & \num{52.00633484} & \num{58.13605453} & \num{12.29567723} & \num{45.75223259} & 3 & \num[scientific-notation=true]{2723.723882} & \num[scientific-notation=true]{0.7562679533} & \num{3.368813786} & \num{0.8945923878} & \num{0.8055553605} & \num{0.1052040819} & \num{45.75153738} & \num{0.105465925}\\ 
\num{70.0000} & \num{60.43766876} & \num{53.76971958} & \num{60.19868169} & \num{15.22579812} & \num{44.84336603} & 2 & \num[scientific-notation=true]{4515.183963} & \num[scientific-notation=true]{1.341088455} & \num{3.257874347} & \num{0.8443177013} & \num{0.7665457294} & \num{0.204358642} & \num{44.84316662} & \num{0.2047613419}\\ 
\num{72.0000} & \num{62.05431099} & \num{55.4091624} & \num{61.82578414} & \num{22.64860065} & \num{39.04338575} & 3 & \num[scientific-notation=true]{5741.428902} & \num[scientific-notation=true]{2.020174655} & \num{3.18881627} & \num{0.8118369511} & \num{0.7426853461} & \num{0.2277546983} & \num{39.04305092} & \num{0.2283174045}\\ 
\num{74.0000} & \num{63.71001032} & \num{56.84227498} & \num{63.49400045} & \num{31.8956343} & \num{31.50220086} & 9 & \num[scientific-notation=true]{6992.466993} & \num[scientific-notation=true]{2.979512442} & \num{3.098565976} & \num{0.7795156993} & \num{0.7191874275} & \num{0.1210069014} & \num{31.50182661} & \num{0.1213568702}\\ 
\num{75.5000} & \num{64.94628468} & \num{58.08242833} & \num{64.73745893} & \num{48.36218678} & \num{16.35312968} & 1 & \num[scientific-notation=true]{10847.95288} & \num[scientific-notation=true]{3.780491839} & \num{3.036835475} & \num{0.7594628082} & \num{0.7045768699} & \num{0.3243103882} & \num{16.35037017} & \num{0.3235235155}\\ 
\num{76.0000} & \num{65.34325729} & \num{58.37855269} & \num{65.12872459} & \num{65.12872459} & - & 1 & - & \num[scientific-notation=true]{4.132206609} & \num{3.015091629} & \num{0.7422054957} & \num{0.6859111505} & - & - & -\\ 
\num{76.5000} & \num{65.76242951} & \num{58.65248156} & \num{65.55937251} & \num{65.55937251} & - & 1 & - & \num[scientific-notation=true]{4.405925784} & \num{2.990994481} & \num{0.7420677234} & \num{0.6912440824} & - & - & -\\ 
\num{77.5000} & \num{66.5865464} & \num{59.53076096} & \num{66.38457139} & \num{66.38457139} & - & 1 & - & \num[scientific-notation=true]{5.05988762} & \num{2.958487154} & \num{0.7286403179} & \num{0.6799624373} & - & - & -\\ 
\num{80.0000} & \num{68.66325003} & \num{61.565736} & \num{68.4658858} & \num{68.4658858} & - & 1 & - & \num[scientific-notation=true]{6.92691067} & \num{2.844695555} & \num{0.691463932} & \num{0.6481610433} & - & - & -\\ 
\hline
\end{tabular}
\end{sidewaystable*}

\begin{sidewaystable*}
\caption{Summary of results for rotating models without angular momentum transport via the ST dynamo.}\label{table:rotnoST}
\centering
\begin{tabular}{ccccccccccccccc} 
\hline\hline             
$M_{\rm i}$ &$M_{\rm He\;dep}$ &$M_{\rm CO,\;He\;dep}$ &$M_{\rm pre\;PPISN/PISN}$ &$M_{\rm ejecta}$ &$M_{\rm CC}$ &\# of pulses &Duration &max KE &$a_{\rm i}$ &$a_{\rm He\;dep}$ &$a_{\rm pre\;PPISN/PISN}$ &$a_{\rm CC}$ &$M_{\rm BH}$ &$a_{\rm BH}$\\ 
$(M_{\odot})$ &$(M_{\odot})$ &$(M_{\odot})$ &$(M_{\odot})$ &$(M_{\odot})$ &$(M_{\odot})$ & &(yr) &$10^{51} [erg]$ & & & & &$(M_{\odot})$ &\\ 
\hline
\num{30.0000} & \num{27.39839602} & \num{23.68757406} & - & - & \num{27.34491066} & 0 & - & - & \num{5.807341127} & \num{3.071151993} & \num{3.018899638} & \num{3.018899638} & \num{22.37731145} & \num{0.9950282185}\\ 
\num{35.0000} & \num{31.69526328} & \num{27.42337917} & - & - & \num{31.63882175} & 0 & - & - & \num{5.299643917} & \num{2.694822567} & \num{2.658873799} & \num{2.658873799} & \num{26.69558257} & \num{0.9927976194}\\ 
\num{40.0000} & \num{35.99867339} & \num{31.25904102} & - & - & \num{35.93415002} & 0 & - & - & \num{4.842120059} & \num{2.416085219} & \num{2.384963179} & \num{2.384963179} & \num{30.74171718} & \num{0.9914386767}\\ 
\num{45.0000} & \num{40.30270527} & \num{35.01112679} & - & - & \num{40.23452207} & 0 & - & - & \num{4.480603714} & \num{2.243489765} & \num{2.220422187} & \num{2.220422187} & \num{34.12579756} & \num{0.98991008}\\ 
\num{47.0000} & \num{41.94645561} & \num{36.56717687} & - & - & \num{41.87069743} & 0 & - & - & \num{4.36729344} & \num{2.125122963} & \num{2.099368204} & \num{2.099368193} & \num{36.47572268} & \num{0.9880213348}\\ 
\num{48.0000} & \num{42.84348452} & \num{37.4778491} & - & - & \num{42.77324571} & 0 & - & - & \num{4.31506298} & \num{2.16280389} & \num{2.144770646} & \num{2.14477066} & \num{36.97073476} & \num{0.9826506525}\\ 
\num{48.5000} & \num{43.26663967} & \num{38.03398711} & \num{43.1941669} & \num{0.1334536901} & \num{43.06071321} & 1 & \num[scientific-notation=true]{0.000703596801} & \num[scientific-notation=true]{0.006809779785} & \num{4.279415518} & \num{2.144129627} & \num{2.125002031} & \num{2.112243553} & \num{37.49964766} & \num{0.9880080513}\\ 
\num{49.0000} & \num{43.71378388} & \num{38.32584843} & \num{43.64188977} & \num{0.2796678404} & \num{43.36222193} & 1 & \num[scientific-notation=true]{0.001067952835} & \num[scientific-notation=true]{0.02688446352} & \num{4.23359568} & \num{2.134283514} & \num{2.116759872} & \num{2.096340352} & \num{37.65069739} & \num{0.9880782754}\\ 
\num{50.0000} & \num{44.55146437} & \num{38.9532359} & \num{44.47980418} & \num{0.2240286373} & \num{44.25577554} & 1 & \num[scientific-notation=true]{0.000811531092} & \num[scientific-notation=true]{0.01168915119} & \num{4.193741077} & \num{2.11150661} & \num{2.096028724} & \num{2.079327209} & \num{38.73913896} & \num{0.9870247924}\\ 
\num{55.0000} & \num{48.79934554} & \num{43.01880946} & \num{48.72078691} & \num{0.3653508166} & \num{48.3554361} & 4 & \num[scientific-notation=true]{0.05959632073} & \num[scientific-notation=true]{0.01372323607} & \num{3.900184065} & \num{1.983047202} & \num{1.970908094} & \num{1.949780204} & \num{43.39008997} & \num{0.9828528792}\\ 
\num{62.5000} & \num{55.01858314} & \num{48.62656499} & \num{54.92342907} & \num{2.222490424} & \num{52.70093865} & 5 & \num[scientific-notation=true]{0.3027699402} & \num[scientific-notation=true]{0.0880625044} & \num{3.566375719} & \num{1.770716195} & \num{1.757868764} & \num{1.666978706} & \num{46.86116548} & \num{0.9779954311}\\ 
\num{70.0000} & \num{61.2959271} & \num{54.43745425} & \num{61.1865688} & \num{8.048999308} & \num{53.13756949} & 5 & \num[scientific-notation=true]{2.323549998} & \num[scientific-notation=true]{0.2282665167} & \num{3.257077128} & \num{1.64246616} & \num{1.631442556} & \num{1.379071398} & \num{48.92057696} & \num{0.9555059834}\\ 
\num{72.5000} & \num{63.42169987} & \num{56.30590456} & \num{63.3132853} & \num{9.137682525} & \num{54.1750397} & 5 & \num[scientific-notation=true]{36.55214653} & \num[scientific-notation=true]{0.3139838485} & \num{3.162543815} & \num{1.633487938} & \num{1.625959435} & \num{1.404570205} & \num{49.66973221} & \num{0.960162202}\\ 
\num{75.0000} & \num{65.49080746} & \num{58.22132638} & \num{65.3746441} & \num{7.654103968} & \num{57.71729217} & 3 & \num[scientific-notation=true]{111.0159479} & \num[scientific-notation=true]{0.2795876214} & \num{3.060682067} & \num{1.575514312} & \num{1.566689181} & \num{1.38822038} & \num{52.4316449} & \num{0.9462283259}\\ 
\num{77.5000} & \num{67.61959384} & \num{60.1890704} & \num{67.50113768} & \num{10.8966896} & \num{56.56128979} & 3 & \num[scientific-notation=true]{1095.767743} & \num[scientific-notation=true]{0.4676799511} & \num{2.958496648} & \num{1.55301032} & \num{1.545782467} & \num{1.339872563} & \num{51.62955332} & \num{0.9489045923}\\ 
\num{80.0000} & \num{69.71228439} & \num{62.11670072} & \num{69.58477642} & \num{15.02132949} & \num{54.40974968} & 2 & \num[scientific-notation=true]{4076.809889} & \num[scientific-notation=true]{1.089654349} & \num{2.844704713} & \num{1.495880554} & \num{1.487293971} & \num{1.208045478} & \num{48.98517551} & \num{0.8968115113}\\ 
\num{82.5000} & \num{71.81110697} & \num{64.0146818} & \num{71.68110957} & \num{21.46695518} & \num{49.9809865} & 2 & \num[scientific-notation=true]{6749.133302} & \num[scientific-notation=true]{1.74115715} & \num{2.754030103} & \num{1.469808123} & \num{1.46273078} & \num{1.121609577} & \num{45.23363553} & \num{0.8978664593}\\ 
\num{85.0000} & \num{73.92021823} & \num{65.98416989} & \num{73.78564943} & \num{27.65649079} & \num{45.95019182} & 2 & \num[scientific-notation=true]{6293.843215} & \num[scientific-notation=true]{2.873729499} & \num{2.666435599} & \num{1.436817835} & \num{1.429992518} & \num{0.9550179694} & \num{41.80532906} & \num{0.8209600511}\\ 
\num{88.2500} & \num{76.58590376} & \num{68.41444571} & \num{76.44570929} & \num{53.63670632} & \num{22.76707263} & 2 & \num[scientific-notation=true]{9932.9834} & \num[scientific-notation=true]{4.546758413} & \num{2.595650549} & \num{1.409184427} & \num{1.40286262} & \num{0.8886053065} & \num{22.01441016} & \num{0.8438832101}\\ 
\num{88.5000} & \num{76.81950452} & \num{68.70116512} & \num{76.67763366} & \num{76.67763366} & - & 1 & - & \num[scientific-notation=true]{4.796166608} & \num{2.573639434} & \num{1.400211784} & \num{1.393786162} & - & - & -\\ 
\num{90.0000} & \num{77.92416739} & \num{69.79673732} & \num{77.77883818} & \num{77.77883818} & - & 1 & - & \num[scientific-notation=true]{5.423466249} & \num{2.614406525} & \num{1.404305018} & \num{1.397687771} & - & - & -\\ 
\num{95.0000} & \num{81.77657178} & \num{73.37959449} & \num{81.61945493} & \num{81.61945493} & - & 1 & - & \num[scientific-notation=true]{8.616296238} & \num{2.624888176} & \num{1.390262109} & \num{1.383494499} & - & - & -\\ 
\num{100.0000} & \num{85.68630916} & \num{76.84979243} & \num{85.51716557} & \num{85.51716557} & - & 1 & - & \num[scientific-notation=true]{11.71137784} & \num{2.591404574} & \num{1.366829047} & \num{1.360339902} & - & - & -\\ 
\hline
\end{tabular}
\end{sidewaystable*}

\section{Spin posterior of the primary BH of GW170729}\label{app:gw170729}
\begin{figure}
   \centering
   \includegraphics[width=\columnwidth]{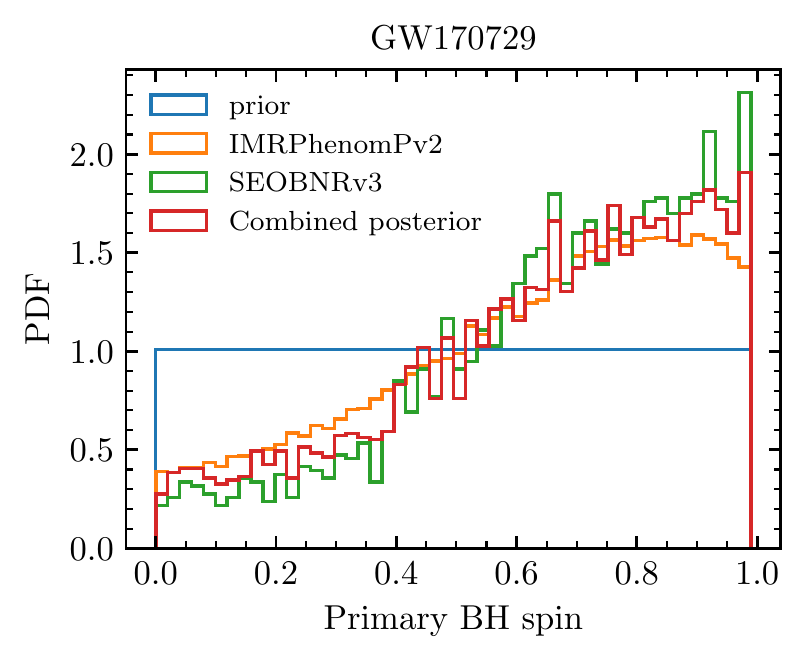}
   \caption{Posterior distribution for the spin of the more massive BH in GW170729 using different waveform models, compared to the flat prior distribution used. The "Combined posterior" is obtained by using an equal number of random samples from each of the other two.}
              \label{fig:spin}%
\end{figure}

Although the first catalogue of gravitational wave transients \citep{GWTC1} does not provide confidence intervals for individual BH spins, the posterior samples they computed are openly available and can be used to obtain this information\footnote{\url{https://doi.org/10.7935/KSX7-QQ51}}. These posteriors are computed using two waveform models, IMRPhenonPv2 \citep{Hannam+2014} and SEOBNRv3 \citep{Bohe+2017}, as well as a set that combines an equal number of samples from both waveform models. Fig. \ref{fig:spin} shows the posterior distributions compared to the prior used for parameter estimation, which corresponds to a flat distribution in spin between $0\leq a\leq 0.99$. From the combined distribution we find a median value with a $90\%$ confidence interval of $a=0.69^{+0.27}_{-0.55}$.


\begin{thebibliography}{79}
\expandafter\ifx\csname natexlab\endcsname\relax\def\natexlab#1{#1}\fi

\bibitem[{{Abbott} {et~al.}(2018{\natexlab{a}}){Abbott}, {Abbott}, {Abbott},
  {Abraham}, \& et~al.}]{LIGOpop}
{Abbott}, B.~P., {Abbott}, R., {Abbott}, T.~D., {Abraham}, S., \& et~al.
  2018{\natexlab{a}}, arXiv e-prints, arXiv:1811.12940

\bibitem[{{Abbott} {et~al.}(2018{\natexlab{b}}){Abbott}, {Abbott}, {Abbott},
  {Abraham}, \& et~al.}]{GWTC1}
{Abbott}, B.~P., {Abbott}, R., {Abbott}, T.~D., {Abraham}, S., \& et~al.
  2018{\natexlab{b}}, arXiv e-prints, arXiv:1811.12907

\bibitem[{{Aguilera-Dena} {et~al.}(2018){Aguilera-Dena}, {Langer}, {Moriya}, \&
  {Schootemeijer}}]{Aguilera-Dena+2018}
{Aguilera-Dena}, D.~R., {Langer}, N., {Moriya}, T.~J., \& {Schootemeijer}, A.
  2018, \apj, 858, 115

\bibitem[{{Angulo} {et~al.}(1999){Angulo}, {Arnould}, {Rayet}, {Descouvemont},
  {Baye}, {Leclercq-Willain}, {Coc}, {Barhoumi}, {Aguer}, {Rolfs}, {Kunz},
  {Hammer}, {Mayer}, {Paradellis}, {Kossionides}, {Chronidou}, {Spyrou},
  {degl'Innocenti}, {Fiorentini}, {Ricci}, {Zavatarelli}, {Providencia},
  {Wolters}, {Soares}, {Grama}, {Rahighi}, {Shotter}, \& {Lamehi
  Rachti}}]{Angulo+1999}
{Angulo}, C., {Arnould}, M., {Rayet}, M., {et~al.} 1999, Nuclear Physics A,
  656, 3

\bibitem[{{Arcavi} {et~al.}(2017){Arcavi}, {Howell}, {Kasen}, {Bildsten},
  {Hosseinzadeh}, {McCully}, {Wong}, {Katz}, {Gal-Yam}, {Sollerman}, {Taddia},
  {Leloudas}, {Fremling}, {Nugent}, {Horesh}, {Mooley}, {Rumsey}, {Cenko},
  {Graham}, {Perley}, {Nakar}, {Shaviv}, {Bromberg}, {Shen}, {Ofek}, {Cao},
  {Wang}, {Huang}, {Rui}, {Zhang}, {Li}, {Li}, {Zhang}, {Valenti}, {Guevel},
  {Shappee}, {Kochanek}, {Holoien}, {Filippenko}, {Fender}, {Nyholm}, {Yaron},
  {Kasliwal}, {Sullivan}, {Blagorodnova}, {Walters}, {Lunnan}, {Khazov},
  {Andreoni}, {Laher}, {Konidaris}, {Wozniak}, \& {Bue}}]{Arcavi+2017}
{Arcavi}, I., {Howell}, D.~A., {Kasen}, D., {et~al.} 2017, \nat, 551, 210

\bibitem[{{Asplund} {et~al.}(2009){Asplund}, {Grevesse}, {Sauval}, \&
  {Scott}}]{Asplund+2009}
{Asplund}, M., {Grevesse}, N., {Sauval}, A.~J., \& {Scott}, P. 2009, \araa, 47,
  481

\bibitem[{{Batta} \& {Ramirez-Ruiz}(2019)}]{BattaRamirezruiz2019}
{Batta}, A. \& {Ramirez-Ruiz}, E. 2019, arXiv e-prints, arXiv:1904.04835

\bibitem[{{Belczynski} {et~al.}(2014){Belczynski}, {Buonanno}, {Cantiello},
  {Fryer}, {Holz}, {Mandel}, {Miller}, \& {Walczak}}]{Belczynski+2014}
{Belczynski}, K., {Buonanno}, A., {Cantiello}, M., {et~al.} 2014, \apj, 789,
  120

\bibitem[{{Belczynski} {et~al.}(2016){Belczynski}, {Heger}, {Gladysz},
  {Ruiter}, {Woosley}, {Wiktorowicz}, {Chen}, {Bulik}, {O'Shaughnessy}, {Holz},
  {Fryer}, \& {Berti}}]{Belczynski+2016b}
{Belczynski}, K., {Heger}, A., {Gladysz}, W., {et~al.} 2016, \aap, 594, A97

\bibitem[{{Boh{\'e}} {et~al.}(2017){Boh{\'e}}, {Shao}, {Taracchini},
  {Buonanno}, {Babak}, {Harry}, {Hinder}, {Ossokine}, {P{\"u}rrer}, {Raymond},
  {Chu}, {Fong}, {Kumar}, {Pfeiffer}, {Boyle}, {Hemberger}, {Kidder},
  {Lovelace}, {Scheel}, \& {Szil{\'a}gyi}}]{Bohe+2017}
{Boh{\'e}}, A., {Shao}, L., {Taracchini}, A., {et~al.} 2017, \prd, 95, 044028

\bibitem[{{B{\"o}hm-Vitense}(1958)}]{Bohm-Vitense1958}
{B{\"o}hm-Vitense}, E. 1958, Zeitschrift f\"ur Astrophysik, 46, 108

\bibitem[{{Campana} {et~al.}(2006){Campana}, {Mangano}, {Blustin}, {Brown},
  {Burrows}, {Chincarini}, {Cummings}, {Cusumano}, {Della Valle}, {Malesani},
  {M{\'e}sz{\'a}ros}, {Nousek}, {Page}, {Sakamoto}, {Waxman}, {Zhang}, {Dai},
  {Gehrels}, {Immler}, {Marshall}, {Mason}, {Moretti}, {O'Brien}, {Osborne},
  {Page}, {Romano}, {Roming}, {Tagliaferri}, {Cominsky}, {Giommi}, {Godet},
  {Kennea}, {Krimm}, {Angelini}, {Barthelmy}, {Boyd}, {Palmer}, {Wells}, \&
  {White}}]{Campana+2006}
{Campana}, S., {Mangano}, V., {Blustin}, A.~J., {et~al.} 2006, \nat, 442, 1008

\bibitem[{{Caughlan} \& {Fowler}(1988)}]{CaughlanFowler1988}
{Caughlan}, G.~R. \& {Fowler}, W.~A. 1988, Atomic Data and Nuclear Data Tables,
  40, 283

\bibitem[{{Chaboyer} \& {Zahn}(1992)}]{ChaboyerZahn1992}
{Chaboyer}, B. \& {Zahn}, J.-P. 1992, \aap, 253, 173

\bibitem[{{Chatzopoulos} {et~al.}(2013){Chatzopoulos}, {Wheeler}, \&
  {Couch}}]{Chatzopoulos+2013}
{Chatzopoulos}, E., {Wheeler}, J.~C., \& {Couch}, S.~M. 2013, \apj, 776, 129

\bibitem[{{Cox} \& {Giuli}(1968)}]{CoxGiuli1968}
{Cox}, J.~P. \& {Giuli}, R.~T. 1968, {Principles of stellar structure } (Gordon
  \& Breach)

\bibitem[{{Denissenkov} \& {Pinsonneault}(2007)}]{DenissenkovPinsonneault2007}
{Denissenkov}, P.~A. \& {Pinsonneault}, M. 2007, \apj, 655, 1157

\bibitem[{{Di Carlo} {et~al.}(2019){Di Carlo}, {Giacobbo}, {Mapelli},
  {Pasquato}, {Spera}, {Wang}, \& {Haardt}}]{diCarlo+2019}
{Di Carlo}, U.~N., {Giacobbo}, N., {Mapelli}, M., {et~al.} 2019, \mnras, 487,
  2947

\bibitem[{{Endal} \& {Sofia}(1976)}]{EndalSofia1976}
{Endal}, A.~S. \& {Sofia}, S. 1976, \apj, 210, 184

\bibitem[{{Farmer} {et~al.}(2020){Farmer}, {Renzo}, {de Mink}, {Fishbach}, \&
  {Justham}}]{Farmer+2020}
{Farmer}, R., {Renzo}, M., {de Mink}, S., {Fishbach}, M., \& {Justham}, S.
  2020, arXiv e-prints, arXiv:2006.06678

\bibitem[{{Farmer} {et~al.}(2019){Farmer}, {Renzo}, {de Mink}, {Marchant}, \&
  {Justham}}]{Farmer+2019}
{Farmer}, R., {Renzo}, M., {de Mink}, S.~E., {Marchant}, P., \& {Justham}, S.
  2019, \apj, 887, 53

\bibitem[{{Farr} {et~al.}(2019){Farr}, {Fishbach}, {Ye}, \& {Holz}}]{Farr+2019}
{Farr}, W.~M., {Fishbach}, M., {Ye}, J., \& {Holz}, D.~E. 2019, \apjl, 883, L42

\bibitem[{{Ferguson} {et~al.}(2005){Ferguson}, {Alexander}, {Allard}, {Barman},
  {Bodnarik}, {Hauschildt}, {Heffner-Wong}, \& {Tamanai}}]{Ferguson+2005}
{Ferguson}, J.~W., {Alexander}, D.~R., {Allard}, F., {et~al.} 2005, \apj, 623,
  585

\bibitem[{{Fishbach} \& {Holz}(2017)}]{FishbachHolz2017}
{Fishbach}, M. \& {Holz}, D.~E. 2017, \apjl, 851, L25

\bibitem[{{Fowler} \& {Hoyle}(1964)}]{FowlerHoyle1964}
{Fowler}, W.~A. \& {Hoyle}, F. 1964, \apjs, 9, 201

\bibitem[{{Fraley}(1968)}]{Fraley1968}
{Fraley}, G.~S. 1968, \apss, 2, 96

\bibitem[{{Fuller} {et~al.}(2019){Fuller}, {Piro}, \& {Jermyn}}]{Fuller+2019}
{Fuller}, J., {Piro}, A.~L., \& {Jermyn}, A.~S. 2019, \mnras, 485, 3661

\bibitem[{{Gal-Yam} {et~al.}(2009){Gal-Yam}, {Mazzali}, {Ofek}, {Nugent},
  {Kulkarni}, {Kasliwal}, {Quimby}, {Filippenko}, {Cenko}, {Chornock},
  {Waldman}, {Kasen}, {Sullivan}, {Beshore}, {Drake}, {Thomas}, {Bloom},
  {Poznanski}, {Miller}, {Foley}, {Silverman}, {Arcavi}, {Ellis}, \&
  {Deng}}]{Gal-Yam+2009}
{Gal-Yam}, A., {Mazzali}, P., {Ofek}, E.~O., {et~al.} 2009, \nat, 462, 624

\bibitem[{{Galama} {et~al.}(1998){Galama}, {Vreeswijk}, {van Paradijs},
  {Kouveliotou}, {Augusteijn}, {B{\"o}hnhardt}, {Brewer}, {Doublier},
  {Gonzalez}, {Leibundgut}, {Lidman}, {Hainaut}, {Patat}, {Heise}, {in't Zand},
  {Hurley}, {Groot}, {Strom}, {Mazzali}, {Iwamoto}, {Nomoto}, {Umeda},
  {Nakamura}, {Young}, {Suzuki}, {Shigeyama}, {Koshut}, {Kippen}, {Robinson},
  {de Wildt}, {Wijers}, {Tanvir}, {Greiner}, {Pian}, {Palazzi}, {Frontera},
  {Masetti}, {Nicastro}, {Feroci}, {Costa}, {Piro}, {Peterson}, {Tinney},
  {Boyle}, {Cannon}, {Stathakis}, {Sadler}, {Begam}, \& {Ianna}}]{Galama+1998}
{Galama}, T.~J., {Vreeswijk}, P.~M., {van Paradijs}, J., {et~al.} 1998, \nat,
  395, 670

\bibitem[{{Gerosa} \& {Berti}(2017)}]{GerosaBerti2017}
{Gerosa}, D. \& {Berti}, E. 2017, \prd, 95, 124046

\bibitem[{{Glatzel} {et~al.}(1985){Glatzel}, {Fricke}, \& {El
  Eid}}]{Glatzel+1985}
{Glatzel}, W., {Fricke}, K.~J., \& {El Eid}, M.~F. 1985, \aap, 149, 413

\bibitem[{{Greiner} {et~al.}(2015){Greiner}, {Mazzali}, {Kann}, {Kr{\"u}hler},
  {Pian}, {Prentice}, {Olivares E.}, {Rossi}, {Klose}, {Taubenberger}, {Knust},
  {Afonso}, {Ashall}, {Bolmer}, {Delvaux}, {Diehl}, {Elliott}, {Filgas},
  {Fynbo}, {Graham}, {Guelbenzu}, {Kobayashi}, {Leloudas}, {Savaglio},
  {Schady}, {Schmidl}, {Schweyer}, {Sudilovsky}, {Tanga}, {Updike}, {van
  Eerten}, \& {Varela}}]{greiner2015}
{Greiner}, J., {Mazzali}, P.~A., {Kann}, D.~A., {et~al.} 2015, \nat, 523, 189

\bibitem[{{Hamann} {et~al.}(1995){Hamann}, {Koesterke}, \&
  {Wessolowski}}]{Hamann+1995}
{Hamann}, W.-R., {Koesterke}, L., \& {Wessolowski}, U. 1995, \aap, 299, 151

\bibitem[{{Hannam} {et~al.}(2013){Hannam}, {Brown}, {Fairhurst}, {Fryer}, \&
  {Harry}}]{Hannam+2013}
{Hannam}, M., {Brown}, D.~A., {Fairhurst}, S., {Fryer}, C.~L., \& {Harry},
  I.~W. 2013, \apjl, 766, L14

\bibitem[{{Hannam} {et~al.}(2014){Hannam}, {Schmidt}, {Boh{\'e}}, {Haegel},
  {Husa}, {Ohme}, {Pratten}, \& {P{\"u}rrer}}]{Hannam+2014}
{Hannam}, M., {Schmidt}, P., {Boh{\'e}}, A., {et~al.} 2014, \prl, 113, 151101

\bibitem[{{Heger} \& {Langer}(2000)}]{HegerLanger2000}
{Heger}, A. \& {Langer}, N. 2000, \apj, 544, 1016

\bibitem[{{Heger} {et~al.}(2000){Heger}, {Langer}, \& {Woosley}}]{Heger+2000}
{Heger}, A., {Langer}, N., \& {Woosley}, S.~E. 2000, \apj, 528, 368

\bibitem[{{Heger} \& {Woosley}(2002)}]{HegerWoosley2002}
{Heger}, A. \& {Woosley}, S.~E. 2002, \apj, 567, 532

\bibitem[{{Herwig}(2000)}]{Herwig2000}
{Herwig}, F. 2000, \aap, 360, 952

\bibitem[{{Iglesias} \& {Rogers}(1996)}]{IglesiasRogers1996}
{Iglesias}, C.~A. \& {Rogers}, F.~J. 1996, \apj, 464, 943

\bibitem[{{Kimball} {et~al.}(2020){Kimball}, {Berry}, \&
  {Kalogera}}]{Kimball+2020}
{Kimball}, C., {Berry}, C., \& {Kalogera}, V. 2020, Research Notes of the
  American Astronomical Society, 4, 2

\bibitem[{{Langer}(1997)}]{Langer1997}
{Langer}, N. 1997, in Astronomical Society of the Pacific Conference Series,
  Vol. 120, Luminous Blue Variables: Massive Stars in Transition, ed. A.~{Nota}
  \& H.~{Lamers}, 83

\bibitem[{{Langer} {et~al.}(1983){Langer}, {Fricke}, \&
  {Sugimoto}}]{Langer+1983}
{Langer}, N., {Fricke}, K.~J., \& {Sugimoto}, D. 1983, \aap, 126, 207

\bibitem[{{Levan} {et~al.}(2014){Levan}, {Tanvir}, {Starling}, {Wiersema},
  {Page}, {Perley}, {Schulze}, {Wynn}, {Chornock}, {Hjorth}, {Cenko},
  {Fruchter}, {O'Brien}, {Brown}, {Tunnicliffe}, {Malesani}, {Jakobsson},
  {Watson}, {Berger}, {Bersier}, {Cobb}, {Covino}, {Cucchiara}, {de Ugarte
  Postigo}, {Fox}, {Gal-Yam}, {Goldoni}, {Gorosabel}, {Kaper}, {Kr{\"u}hler},
  {Karjalainen}, {Osborne}, {Pian}, {S{\'a}nchez-Ram{\'\i}rez}, {Schmidt},
  {Skillen}, {Tagliaferri}, {Th{\"o}ne}, {Vaduvescu}, {Wijers}, \&
  {Zauderer}}]{levan2014ulgrb}
{Levan}, A.~J., {Tanvir}, N.~R., {Starling}, R.~L.~C., {et~al.} 2014, \apj,
  781, 13

\bibitem[{{Lunnan} {et~al.}(2018){Lunnan}, {Fransson}, {Vreeswijk}, {Woosley},
  {Leloudas}, {Perley}, {Quimby}, {Yan}, {Blagorodnova}, {Bue}, {Cenko}, {De
  Cia}, {Cook}, {Fremling}, {Gatkine}, {Gal-Yam}, {Kasliwal}, {Kulkarni},
  {Masci}, {Nugent}, {Nyholm}, {Rubin}, {Suzuki}, \& {Wozniak}}]{Lunnan+2018}
{Lunnan}, R., {Fransson}, C., {Vreeswijk}, P.~M., {et~al.} 2018, Nature
  Astronomy [\eprint[arXiv]{1808.04887}]

\bibitem[{{MacFadyen} \& {Woosley}(1999)}]{MacfadyenWoosley1999}
{MacFadyen}, A.~I. \& {Woosley}, S.~E. 1999, \apj, 524, 262

\bibitem[{{Mandel} {et~al.}(2019){Mandel}, {Farr}, \& {Gair}}]{Mandel+2019}
{Mandel}, I., {Farr}, W.~M., \& {Gair}, J.~R. 2019, \mnras, 486, 1086

\bibitem[{{Mapelli} {et~al.}(2020){Mapelli}, {Spera}, {Montanari}, {Limongi},
  {Chieffi}, {Giacobbo}, {Bressan}, \& {Bouffanais}}]{Mapelli+2020}
{Mapelli}, M., {Spera}, M., {Montanari}, E., {et~al.} 2020, \apj, 888, 76

\bibitem[{{Marchant} {et~al.}(2016){Marchant}, {Langer}, {Podsiadlowski},
  {Tauris}, \& {Moriya}}]{Marchant+2016}
{Marchant}, P., {Langer}, N., {Podsiadlowski}, P., {Tauris}, T.~M., \&
  {Moriya}, T.~J. 2016, \aap, 588, A50

\bibitem[{{Marchant} {et~al.}(2019){Marchant}, {Renzo}, {Farmer}, {Pappas},
  {Taam}, {de Mink}, \& {Kalogera}}]{Marchant+2019}
{Marchant}, P., {Renzo}, M., {Farmer}, R., {et~al.} 2019, \apj, 882, 36

\bibitem[{{Moriya} {et~al.}(2020){Moriya}, {Marchant}, \& {Blinnikov}}]{mmb20}
{Moriya}, T.~J., {Marchant}, P., \& {Blinnikov}, S.~I. 2020, \aap, submitted

\bibitem[{{Nieuwenhuijzen} \& {de Jager}(1990)}]{NieuwenhuijzendeJager1990}
{Nieuwenhuijzen}, H. \& {de Jager}, C. 1990, \aap, 231, 134

\bibitem[{{Paxton} {et~al.}(2011){Paxton}, {Bildsten}, {Dotter}, {Herwig},
  {Lesaffre}, \& {Timmes}}]{Paxton+2011}
{Paxton}, B., {Bildsten}, L., {Dotter}, A., {et~al.} 2011, \apjs, 192, 3

\bibitem[{{Paxton} {et~al.}(2013){Paxton}, {Cantiello}, {Arras}, {Bildsten},
  {Brown}, {Dotter}, {Mankovich}, {Montgomery}, {Stello}, {Timmes}, \&
  {Townsend}}]{Paxton+2013}
{Paxton}, B., {Cantiello}, M., {Arras}, P., {et~al.} 2013, \apjs, 208, 4

\bibitem[{{Paxton} {et~al.}(2015){Paxton}, {Marchant}, {Schwab}, {Bauer},
  {Bildsten}, {Cantiello}, {Dessart}, {Farmer}, {Hu}, {Langer}, {Townsend},
  {Townsley}, \& {Timmes}}]{Paxton+2015}
{Paxton}, B., {Marchant}, P., {Schwab}, J., {et~al.} 2015, \apjs, 220, 15

\bibitem[{{Paxton} {et~al.}(2018){Paxton}, {Schwab}, {Bauer}, {Bildsten},
  {Blinnikov}, {Duffell}, {Farmer}, {Goldberg}, {Marchant}, {Sorokina},
  {Thoul}, {Townsend}, \& {Timmes}}]{Paxton+2018}
{Paxton}, B., {Schwab}, J., {Bauer}, E.~B., {et~al.} 2018, \apjs, 234, 34

\bibitem[{{Paxton} {et~al.}(2019){Paxton}, {Smolec}, {Schwab}, {Gautschy},
  {Bildsten}, {Cantiello}, {Dotter}, {Farmer}, {Goldberg}, {Jermyn}, {Kanbur},
  {Marchant}, {Thoul}, {Townsend}, {Wolf}, {Zhang}, \& {Timmes}}]{Paxton+2019}
{Paxton}, B., {Smolec}, R., {Schwab}, J., {et~al.} 2019, \apjs, 243, 10

\bibitem[{{Potekhin} \& {Chabrier}(2010)}]{PotekhinChabrier2010}
{Potekhin}, A.~Y. \& {Chabrier}, G. 2010, Contributions to Plasma Physics, 50,
  82

\bibitem[{{Qin} {et~al.}(2019){Qin}, {Marchant}, {Fragos}, {Meynet}, \&
  {Kalogera}}]{Qin+2019}
{Qin}, Y., {Marchant}, P., {Fragos}, T., {Meynet}, G., \& {Kalogera}, V. 2019,
  \apjl, 870, L18

\bibitem[{{Quataert} \& {Kasen}(2012)}]{quataert2012}
{Quataert}, E. \& {Kasen}, D. 2012, \mnras, 419, L1

\bibitem[{{Rakavy} \& {Shaviv}(1967)}]{RakaviShaviv1967}
{Rakavy}, G. \& {Shaviv}, G. 1967, \apj, 148, 803

\bibitem[{{Renzo} {et~al.}(2020){Renzo}, {Farmer}, {Justham}, {de Mink},
  {G{\"o}tberg}, \& {Marchant}}]{Renzo+2020}
{Renzo}, M., {Farmer}, R.~J., {Justham}, S., {et~al.} 2020, \mnras, 493, 4333

\bibitem[{{Rogers} \& {Nayfonov}(2002)}]{RogersNayfonov2002}
{Rogers}, F.~J. \& {Nayfonov}, A. 2002, \apj, 576, 1064

\bibitem[{{Saumon} {et~al.}(1995){Saumon}, {Chabrier}, \& {van
  Horn}}]{Saumon+1995}
{Saumon}, D., {Chabrier}, G., \& {van Horn}, H.~M. 1995, \apjs, 99, 713

\bibitem[{{Spera} \& {Mapelli}(2017)}]{SperaMapelli2017}
{Spera}, M. \& {Mapelli}, M. 2017, \mnras, 470, 4739

\bibitem[{{Spruit}(1999)}]{Spruit1999}
{Spruit}, H.~C. 1999, \aap, 349, 189

\bibitem[{{Spruit}(2002)}]{Spruit2002}
{Spruit}, H.~C. 2002, \aap, 381, 923

\bibitem[{{Stothers}(1999)}]{Stothers1999}
{Stothers}, R.~B. 1999, \mnras, 305, 365

\bibitem[{{Takahashi}(2018)}]{Takahashi2018}
{Takahashi}, K. 2018, \apj, 863, 153

\bibitem[{{Terreran} {et~al.}(2017){Terreran}, {Pumo}, {Chen}, {Moriya},
  {Taddia}, {Dessart}, {Zampieri}, {Smartt}, {Benetti}, {Inserra},
  {Cappellaro}, {Nicholl}, {Fraser}, {Wyrzykowski}, {Udalski}, {Howell},
  {McCully}, {Valenti}, {Dimitriadis}, {Maguire}, {Sullivan}, {Smith}, {Yaron},
  {Young}, {Anderson}, {Della Valle}, {Elias-Rosa}, {Gal-Yam}, {Jerkstrand},
  {Kankare}, {Pastorello}, {Sollerman}, {Turatto}, {Kostrzewa-Rutkowska},
  {Koz{\l}owski}, {Mr{\'o}z}, {Pawlak}, {Pietrukowicz}, {Poleski}, {Skowron},
  {Skowron}, {Soszy{\'n}ski}, {Szyma{\'n}ski}, \& {Ulaczyk}}]{Terreran+2017}
{Terreran}, G., {Pumo}, M.~L., {Chen}, T.-W., {et~al.} 2017, Nature Astronomy,
  1, 713

\bibitem[{{Timmes} \& {Swesty}(2000)}]{TimmesSwesty2000}
{Timmes}, F.~X. \& {Swesty}, F.~D. 2000, \apjs, 126, 501

\bibitem[{{Toro} {et~al.}(1994){Toro}, {Spruce}, \& {Speares}}]{Toro+1994}
{Toro}, E.~F., {Spruce}, M., \& {Speares}, W. 1994, Shock Waves, 4, 25

\bibitem[{{van Son} {et~al.}(2020){van Son}, {de Mink}, {Broekgaarden},
  {Renzo}, {Justham}, {Laplace}, {Moran-Fraile}, {Hendriks}, \&
  {Farmer}}]{vanSon+2020}
{van Son}, L.~A.~C., {de Mink}, S.~E., {Broekgaarden}, F.~S., {et~al.} 2020,
  arXiv e-prints, arXiv:2004.05187

\bibitem[{{Vink} {et~al.}(2001){Vink}, {de Koter}, \& {Lamers}}]{Vink+2001}
{Vink}, J.~S., {de Koter}, A., \& {Lamers}, H.~J.~G.~L.~M. 2001, \aap, 369, 574

\bibitem[{{Woosley}(1993)}]{Woosley1993}
{Woosley}, S.~E. 1993, \apj, 405, 273

\bibitem[{{Woosley}(2017)}]{Woosley2017}
{Woosley}, S.~E. 2017, \apj, 836, 244

\bibitem[{{Yoshida} {et~al.}(2016){Yoshida}, {Umeda}, {Maeda}, \&
  {Ishii}}]{Yoshida+2016}
{Yoshida}, T., {Umeda}, H., {Maeda}, K., \& {Ishii}, T. 2016, \mnras, 457, 351

\bibitem[{{Zahn} {et~al.}(2007){Zahn}, {Brun}, \& {Mathis}}]{Zahn+2007}
{Zahn}, J.~P., {Brun}, A.~S., \& {Mathis}, S. 2007, \aap, 474, 145

\bibitem[{{Zevin} {et~al.}(2020){Zevin}, {Spera}, {Berry}, \&
  {Kalogera}}]{Zevin+2020}
{Zevin}, M., {Spera}, M., {Berry}, C. P.~L., \& {Kalogera}, V. 2020, arXiv
  e-prints, arXiv:2006.14573

\end{thebibliography}
\end{document}